\documentclass[11pt,twoside]{article}

\usepackage{baaa}

\usepackage{graphicx}
\usepackage{subfigure}
\usepackage{amssymb}
\usepackage[spanish,activeacute,english]{babel}
\usepackage[latin1]{inputenc}
\usepackage{verbatim}
\usepackage{amsmath}
\usepackage{amsfonts}
\usepackage{amssymb}
\usepackage[colorlinks=true,dvips]{hyperref}
\usepackage{natbib}

\begin{document}
%%%%%%%%%%%%%%%%%%%%%%%%%%%%%%%%%%%%%%%%%%%%%%%%%%%%%%%%%%%%%%%%%%%%%%%%%%
%%%% SELECCIONE EL IDIOMA EN QUE SE ESCRIBE EL ART\'{I}CULO:              %%%%
\myselectenglish
%%%%%%%%%%%%%%%%%%%%%%%%%%%%%%%%%%%%%%%%%%%%%%%%%%%%%%%%%%%%%%%%%%%%%%%%%%
\vskip 1.0cm \markboth{Ocampo M. M. et al.}%
{Primordial power spectrum from an objective collapse
mechanism: The simplest case}

\pagestyle{myheadings}

\vspace*{0.5cm}

\noindent PRESENTACI\'{O}N ORAL

\vskip 0.3cm
\title{Primordial power spectrum from an objective collapse
mechanism: The simplest case}

\author{M.M. Ocampo$^{1,2}$, O. Palermo$^{2}$, G. Le\'on$^{2}$ \& G. R. Bengochea$^{3}$}

\affil{%
(1) Instituto de Astrof\'{\i}sica de La Plata, CONICET-UNLP, Argentina \\
(2) Facultad de Ciencias Astron\'omicas y Geof\'{\i}sicas, Universidad Nacional de La Plata, Argentina \\
  (3) Instituto de Astronom\'{\i}a y F\'{\i}sica del Espacio CONICET-UBA, Argentina \\
}

\begin{abstract}
In this work we analyzed the physical origin of the primordial inhomogeneities during the inflation
era. The proposed framework is based, on the one hand, on semiclassical gravity, in which only the matter fields are quantized and not the spacetime metric. Secondly, we incorporate an objective collapse mechanism based on the Continuous Spontaneous Localization (CSL) model, and we apply it to the wavefunction associated with the inflaton field. This is introduced due to the close relation between cosmology and the so-called ``measurement problem'' in Quantum Mechanics. In particular, in order to break the homogeneity and isotropy of the initial Bunch-Davies vacuum, and thus obtain the inhomogeneities observed today, the theory requires something akin to a
``measurement'' (in the traditional sense of Quantum Mechanics). This is because the linear evolution driven by
Schr\"odinger's equation does not break any initial symmetry. The collapse mechanism given by the CSL model provides a satisfactory mechanism for breaking the initial symmetries of the Bunch-Davies vacuum. The novel aspect in this work is that the constructed CSL model arises from the simplest choices for the collapse parameter and operator. From these considerations, we obtain a primordial spectrum that has the same distinctive features as the standard one, which is consistent with the observations from the Cosmic Microwave Background.
\end{abstract}

\begin{resumen}
En este trabajo se analiz\'o el origen f\'isico de las inhomogeneidades primordiales durante la era inflacionaria. El marco te\'orico propuesto est\'a basado, por un lado, en el contexto de gravedad semicl\'asica donde solamente se cuantizan los campos de materia y no la m\'etrica espacio-temporal. Por otro lado, se incorpora un mecanismo de colapsos objetivos basado en el modelo de Localizaci\'on Continua Espont\'anea (CSL por sus siglas en ingl\'es), el cual es aplicado a la funci\'on de onda asociada al campo inflat\'on. Esto es introducido para atender la relaci\'on cercana entre la cosmolog\'ia y el llamado ``problema de la medici\'on'' en Mec\'anica Cu\'antica. En particular, para romper las simetr\'ias asumidas para el vac\'io de Bunch-Davies inicial y as\'i obtener las inhomogeneidades observadas hoy, la teor\'ia necesita de algo similar a una ``medici\'on'', ya que la evoluci\'on lineal dada por la ecuaci\'on de Schr\"odinger no rompe simetr\'ias iniciales. El colapso dado por el modelo CSL provee un mecanismo satisfactorio de ruptura de las simetr\'ias iniciales del vac\'io de Bunch-Davies. El aspecto novedoso en este trabajo es que el modelo CSL propuesto surge a partir de las elecciones m\'as simples para el par\'ametro y operador de colapso, obteni\'endose un espectro primordial que tiene las mismas caracter\'isticas distintivas del espectro est\'andar, y que es consistente con las observaciones del Fondo C\'osmico de Microondas.
\end{resumen}

\section{Introduction}
\label{S_intro}

The current cosmological model provides a description
of the evolution of the Universe, in which the seeds of
structure originate from quantum fluctuations during an
inflationary stage (Guth, 1981; Mukhanov et al., 1992).
The predictions have been tested with success in observations of the Cosmic Microwave Background (CMB)
(Planck Collaboration et al., 2020).

n this standard approach the primordial power spectrum is calculated quantizing both the metric and matter fields. The Universe is pretended to be in an initial symmetric state, and the usual choice is to adopt
the Bunch-Davies (BD) vacuum. This initial symmetric
state is found to evolve into an asymmetric one, with
the inhomogeneities expected to be found in the matter
and energy distribution today. The mechanism responsible for this transition, in the standard model, is the
fluctuations of the vacuum state of the inflaton field.
In practice, the theoretical prediction is consistent with
the CMB observations, and, in particular, for the large
scale modes for which the Sachs-Wolfe effect is dominant (Sachs \& Wolfe, 1967). This corresponds with
the so-called super-horizon modes, i.e. inhomogeneities
that are not affected by microphysics since they are at
scales bigger than the particle horizon at decoupling.
The power spectrum obtained from standard inflation,
denominated Harrison-Zel'dovich spectrum (Harrison,
1970; Zel'dovich, 1972), is scale invariant and given by
the following expression:
\begin{equation} \label{standardps}
    \mathcal{P}_{\mathcal{R}}(k) \simeq \left ( \frac{H_I^2}{2\pi \dot{\phi}}\right)^2 = \frac{H_I^2}{8\pi^2 \varepsilon M^2_P} ,
\end{equation}

where $H_I$ is the Hubble parameter during inflation, $\phi$ is the scalar inflaton field, $M_P$ is the Planck mass, and $\varepsilon$ is the slow-roll parameter. For a complete derivation it is suggested to check Sriramkumar (2009).

However, the physical origin for the primordial perturbations remains unclear, and in Sect. \ref{sec2.1} we will discuss the standard explanation and why we find it inappropriate.

\section{The measurement problem in Quantum Mechanics}

In Sect. \ref{S_intro} we mentioned the fluctuations of the inflaton field in the vacuum state as the mechanism that originates the primordial inhomogeneities in the Universe. We find this argument difficult to justify and it is closely related to the so-called measurement problem in Quantum Physics (Maudlin, 1995; Okon, 2014; Norsen,
2017).

In Quantum Mechanics the equation that describes the temporal evolution of a system from an initial state is the Schr\"odinger equation: 
\begin{equation}
    i \hbar \frac{\partial}{\partial t} | \psi (t) \rangle = \hat{H} | \psi (t) \rangle ,
\end{equation}
where $\hat{H}$ is the Hamiltonian operator and $\hbar$ is the reduced Planck constant. On the other hand, if an external agent (of ambiguous nature intended by the theory) makes a measurement of a property of the system, the wavefunction will collapse and the state will be an eigenstate of the respective operator. This process is random with the probability given by Born's rule.

The evolution given by the Schr\"odinger equation, by itself, is not capable of breaking initial symmetries\footnote{Given that $[\hat{O},\hat{H}]=0$, with $\hat{O}$ being the symmetry generating operator.} and superpositions of a system. The theory needs the measurement postulate in order to connect the formalism with the observations, and such a mechanism provokes the collapse of the wave function. This leads to the following questions: What is a measurement? Who can perform one? When does it happen? When do we have to use the evolution given by Schr\"odinger equation and when the random process that collapses the wavefunction i.e. the measurement postulate? This is the so called \textit{``measurement problem''}.

In Maudlin (1995) the measurement problem is approached formally, synthesizing it in the incompatibility of the following statements:

\begin{enumerate}
    \item The physical description provided by the state vector is complete. \label{A}
    \item Quantum states always evolve according to the Schr\"odinger equation. \label{B}
    \item Measurements always have definite results. \label{C}
\end{enumerate}

Any alternative to the standard quantum formalism that aspires to address the measurement problem must negate at least one of these points. In our case, we will explore a possibility that negates \ref{B}. Alternative approaches that discard the other two statements can be seen in Bell (1992) for \ref{A}, in the known as de Broglie-Bohm model, and Schlosshauer-Selbach (2007) for \ref{C}, where decoherence is explored.

By negating \ref{B} we need to explore theories in which the collapse of the wavefunction is self induced by some novel mechanism. These models are known as \textit{objective collapse theories}. This approach was first explored in works like Pearle (1976), Diosi (1984, 1987, 1989) and
Penrose (1989) with the aim of obtaining a theory that maintains the successful predictions of Quantum Mechanics and can also describe macroscopic phenomena that do not exhibit superposition of states.

In particular, we will use a version of the model called Continuous Spontaneous Localization (CSL), which proposes an objective collapse of the wavefunction (Pearle, 1976, 1989). The quantum-to-classical transition of the perturbations is naturally explained by the CSL model.

\subsection{The cosmological scenario} \label{sec2.1}

In cosmology the measurement problem is enhanced (Bell, 1981; Hartle, 1993; Perez et al., 2006; Sudarsky,
2011; Landau et al., 2013; Bengochea, 2020). We will briefly describe the issue here.    

At the beginning of the inflation stage, both spacetime and the quantum state of the inflaton field are assumed to be isotropic and homogeneous. Then, as in any quantum system, the expected values and quantum uncertainties of the quantum field in that vacuum state can be calculated. On the other hand, until a ``measurement'' is performed, the system will continue having the initial symmetries if its evolution is dictated by the Schr\"odinger equation, which does not break any symmetry. However, today we see the initial symmetries broken, characterized in the anisotropies of the CMB, and also in the structure formation in the Universe. How do we arrive to an asymmetric state from a symmetric one? Usual Quantum Mechanics needs the measurement postulate in order to break symmetries but, in the early Universe, the same questions exposed before emerge: who can make a measurement that breaks symmetries? How is it made in that scenario?

Standard cosmology tries to answer this problem by invoking the Uncertainty Principle and the quantum vacuum fluctuations as the mechanism generator of the seeds of structure. This would mean that the quantum fluctuations have a real existence in the Universe and that the quantum field have real, random but well-defined values at every time. We find this problematic since quantum fluctuations are uncertainties and, thus, the only thing they can provide is the range of the most probable values (together with Born's rule) if a measurement is made. \textbf{Quantum fluctuations are not equivalent to inhomogeneities.}

\section{Continuous Spontaneous Localization}\label{sec:guia}

The Continuous Spontaneous Localization model consists of a non linear modification of the Schr\"odinger equation, to which a stochastic term is added. This model is a generalization of the previously proposed Quantum Mechanics of Spontaneous Localization (QMSL), or GRW model after its authors (Ghirardi, Rimini and Weber ) (Ghirardi et al., 1986). As the name suggests, in CSL the collapse of the wave function occurs continuously, whereas in GRW the collapse is discrete.

CSL has two main equations and their derivation can be found in Pearle (2012). Since we are interested in introducing the CSL mechanism in a cosmological context, we will use an adaptation of them to this scenario. The first equation is a modification of the Schr\"odinger equation to which a non-linear stochastic term is added. The modified evolution equation is:
\begin{eqnarray} \label{eqcsl1}
    |\Phi,\eta\rangle &=& \hat{\mathcal{T}} \text{exp} \bigg \{ \int_{\tau}^{\eta} d\eta \int d^3x \sqrt{|g|} \bigg [ -i\hat{\mathcal{H}}\nonumber 
    \\ 
    &-&
    \frac{1}{4\lambda} (W(\eta,\mathbf{x})-2\lambda \hat{C}(\eta,\mathbf{x}))^2 \bigg ] \bigg \} |\Phi,\tau\rangle .
\end{eqnarray}
In this expression $\Phi$ represents the wavefunction associated with the quantum state of the inhomogeneous part of the inflaton field, $d\eta d^3x\sqrt{|g|}$ is the 4-volume of the background Friedmann-Lema\^itre-Robertson-Walker (FLRW) metric, $\tau \rightarrow -\infty$ the conformal time at the beginning of inflation and $\hat{\mathcal{H}}$ is the Hamiltonian density of the system such that $\hat{H}=\int d^3x\hat{\mathcal{H}}$. $\hat{C}$ is the collapse operator to which one of its eigenstates will collapse the system. $\lambda$ represents the collapse rate and characterizes the \textit{amplification mechanism} of the theory, that is, the collapse being rare for microscopic systems as expected by standard Quantum Mechanics, and increasing this effect to the level of being notable for systems with enough energy to be considered macroscopic. The choices for these two terms will be discussed in the Sect. \ref{sec4}.  $W$ is a scalar white noise field and its probability rule in this case is:
\begin{equation}
    P(W)dW=\langle\Phi,\eta|\Phi,\eta\rangle \prod_{\eta'=\tau}^{\eta-d\eta}\frac{|g|^{1/4}dW(\eta',\mathbf{x})}{\sqrt{2\pi\lambda/d\eta}} .
\end{equation}

Since we are working with conformal time as the time coordinate, we can see that $\sqrt{|g|}=a^4$, with $a$ the usual scale factor in FLRW geometry.

The random nature of $W$ in (\ref{eqcsl1}) is needed because the collapse toward an eigenstate of the collapse operator needs to be random, in order to replicate the predictions of Quantum Mechanics. Then, different systems starting each one in the same initial state will evolve toward different eigenstates. We are interested in the evolution of this ensemble of systems and, to describe it, we introduce the density matrix:
\begin{equation}
    \hat{\rho} \equiv \int_{-\infty}^{\infty} P(W)dW \frac{|\Phi,\eta\rangle\langle\Phi,\eta|}{\langle\Phi,\eta|\Phi,\eta\rangle} ,
\end{equation}
 and, combining this expression with (\ref{eqcsl1}) the following evolution equation can be found:
 \begin{equation}
     \frac{\partial\hat{\rho}}{\partial\eta} = - i [\hat{H},\hat{\rho}] - \frac{\lambda a^4}{2} \int d^3x [\hat{C},[\hat{C},\hat{\rho}]].
 \end{equation}

\section {Collapse in Inflation} \label{sec4}

\subsection{Semiclassical gravity}

In this work we will employ the semiclassical gravity (SCG) framework, as in previous works like Perez et al., (2006) where the collapse proposal for cosmology was originally introduced. The semiclassical Einstein equations are simply:
\begin{equation}
    G_{\mu\nu} = 8\pi G \langle \hat{T}_{\mu\nu}\rangle ,
\end{equation}
with $G_{\mu\nu}$ the Einstein tensor and $\hat{T}_{\mu\nu}$ the energy-momentum tensor.
In this model only $\hat{T}_{\mu\nu}$ is quantized and its expectation value acts as source of spacetime curvature. SCG presents two main advantages (Le\'on \&
Bengochea, 2021):
\begin{itemize}
    \item It is not needed to justify the ``quantum-to-classical'' transition in the metric as the spacetime is always taken as classical. When including the CSL mechanism the collapse of the wave function is regarded as a physical process happening in time, and then is preferred to admit full spacetime notions.
    \item It facilitates presenting how the perturbations are born from the wavefunction collapse, since the expectation value of the energy-momentum tensor in SCG yields a geometry that will not be homogeneous and isotropic after the collapse has taken place. 
\end{itemize}

Nevertheless, CSL calculations with the standard quantization of both spacetime metric and matter fields have been done, for example in Le\'on \& Bengochea
(2016) and Palermo et al. (2022). For a complete discussion about the advantages and disadvantages of SCG approach the interested reader can check Bengochea et al.
(2020).

\subsection{Collapse parameter and operator}

As mentioned before, $\lambda$ represents the collapse rate. Diosi (1984, 1987, 1989) and Penrose (1996) discuss that the collapse mechanism should be a dynamical process related to gravitational interaction. Taking this into account,  Bengochea et al. (2020) and Le\'on \& Bengochea
(2021) consider that $\lambda$ has to be related to the spacetime curvature, and several parameterizations can be considered involving the different mathematical quantities that emerge from the curvature. The simplest choice that we can do is to use the Ricci scalar and, thus, having the following expression:
\begin{equation}
    \lambda = \lambda_0 \frac{R}{M_P} \simeq \lambda_0 \frac{H_I^2}{M_P},
\end{equation}
where $\lambda_0 \simeq 10^{-17}$s$^{-1}$ is estimated from laboratory experiments and $H_I$ is the quasi constant Hubble parameter in slow-roll inflation.

Although the simplest, this choice for $\lambda$ is novel since in previous works like  Ca\~nate et al. (2013), Le\'on \& Bengochea (2016), Le\'on \& Bengochea (2021) and Palermo
et al. (2022) the parameter was not considered as a constant and needed to be properly adjusted in order to be consistent with observations. On the other hand, more sophisticated parameterizations can potentially lead to better predictions for the primordial spectra.

We turn our attention to the collapse operator $\hat{C}$ in (\ref{eqcsl1}). Following the criterion of choosing the simplest quantity, we will use the field variable itself, this means the collapse operator will be the inhomogeneous part of the inflaton field:
\begin{equation}
    \hat{C}\equiv \delta \hat{\phi}.
\end{equation}

\subsection{Primordial power spectrum}

We will show now the relation between the primordial curvature perturbation and the CSL mechanism. For a complete derivation the reader is invited to see  Le\'on
\& Bengochea (2021) and, in particular for this work,
Ocampo (2023).

Using the perturbed Einstein equations in the longitudinal gauge (Mukhanov et al., 1992; Mukhanov, 2005), and the definition of the slow-roll parameter $\varepsilon$, we can obtain  (Le\'on \& Bengochea, 2021):

\begin{equation}
    \Psi + \mathcal{H}^{-1} \Psi' = \sqrt{\frac{\varepsilon}{2}}\frac{\langle \hat{y}\rangle}{a M_P} ,
\end{equation}
where $\Psi$ is a scalar field corresponding to the scalar perturbations at first order and, in the longitudinal gauge, represents the curvature perturbation. $\mathcal{H}$ is the comoving Hubble parameter and $\hat{y}=a\delta\hat{\phi}$. This expression is exact, with no approximations made, and relates the quantum expectation value of the matter fields and the classical curvature perturbations. Since the matter fields follow the evolution driven by CSL equation, we have derived an expression that relates the CSL mechanism (present in $\langle\hat{y}\rangle$) to the scalar perturbations of the metric (given by $\Psi$).

Next, we introduce the Lukash variable (Lukash,
1980), and using the slow-roll approximation, we can relate it to the last expression as follows:
\begin{equation}
    \mathcal{R} \simeq \frac{1}{\varepsilon} (\Psi + \mathcal{H}^{-1} \Psi' ) = \frac{\langle \hat{y}\rangle}{a M_P\sqrt{2\varepsilon}} .
\end{equation}
The variable $\mathcal{R}$ also represents, in the comoving gauge, the curvature perturbation. The associated scalar power spectrum is defined in Fourier space as:
\begin{equation}
    \overline{\mathcal{R}_{\mathbf{k}}\mathcal{R}^*_{\mathbf{q}}} \equiv \frac{2\pi^2}{k^3} \mathcal{P}_{\mathcal{R}}(k)\delta(\mathbf{k}-\mathbf{q}) ,
\end{equation}
where the bar indicates an ensemble average over the field $\mathcal{R}_\mathbf{k}$. Since $\mathcal{R}$ is a gauge invariant quantity, we can compute it in the Newtonian gauge and then switch to the comoving gauge in order to compute the resultant power spectrum. This will be:

\begin{equation}
    \mathcal{P}_{\mathcal{R}}(k)\delta(\mathbf{k}-\mathbf{q})= \frac{k^3}{4\pi^2\varepsilon M^2_P a^2} \overline{\langle\hat{y}_{\mathbf{k}}\rangle\langle\hat{y}_{\mathbf{q}}\rangle^*}.
\end{equation}

Finally, using the CSL parameter and operator chosen in the previous section, we can obtain the following expression for the scalar power spectrum (Ocampo, 2023): 
\begin{equation}
    P_\mathcal{R}(k) \approx \frac{H_I^2}{8\pi^2\varepsilon M_P^2} \left ( 1 - \frac{4 \lambda_0\pi}{M_P}\right),
\end{equation}
this means, the same standard Harrison-Zel'dovich spectrum plus a factor dependent of $\lambda_0$. This factor, in Planck units, results $\lambda_0\simeq 10^{-61}M_P$ and, thus, we can neglect it recovering (\ref{standardps}).

\section{Conclusions}

In this work we discussed the measurement problem in Quantum Physics and its relevance to the cosmological scenario, in which the lack of an external observer for the early Universe (or even a good definition for a measurement) leads to a problem for breaking the primordial symmetries. We explained why we find the standard approach unsatisfactory due to the inappropriate treatment of quantum uncertainties as physical inhomogeneities.

Then, we introduced the CSL model which incorporates an objective collapse mechanism for the wavefunction and chose the simplest options for the collapse rate parameter $\lambda$ and the collapse operator $\hat{C}$. With these choices we computed the primordial scalar power spectrum, obtaining an equal expression to the traditional one. This means that the CSL mechanism provides a plausible explanation for the quantum origin for the primordial perturbations and, with even the simplest parameterization, can lead to a spectrum consistent with observations.

\references

\texttt{Bell J.S., 1981, Oxford University Press, 611-637}

\texttt{Bell J.S., 1982, Foundations of Physics, 12, 989}

\texttt{Bengochea G.R., 2020, Revista Mexicana de F\'isica E, 17,
263-271}

\texttt{Bengochea G.R., et al., 2020, The European Physical Journal C, 80}

\texttt{Ca\~nate P., Pearle P., Sudarsky D., 2013, Physical Review D,
87}

\texttt{Diosi L., 1984, Phys. Lett. A, 105, 199}

\texttt{Diosi L., 1987, Phys. Lett. A, 120, 377}

\texttt{Diosi L., 1989, Phys. Rev. A, 40, 1165}

\texttt{Ghirardi G.C., Rimini A., Weber T., 1986, PhRvD, 34, 470}

\texttt{Guth A.H., 1981, Phys. Rev. D, 23, 347}

\texttt{Harrison E.R., 1970, Phys. Rev. D, 1, 2726}

\texttt{Hartle J.B., 1993, arXiv e-prints, gr-qc/9304006}

\texttt{Landau S., Le\'on G., Sudarsky D., 2013, Phys. Rev. D, 88,
023526}

\texttt{Le\'on G., Bengochea G.R., 2016, European Physical Journal
C, 76, 29}

\texttt{Le\'on G., Bengochea G.R., 2021, European Physical Journal
C, 81, 1055}

\texttt{Lukash V.N., 1980, Sov. Phys. JETP, 52, 807}

\texttt{Maudlin T., 1995, Topoi, 14, 7}

\texttt{Mukhanov V., 2005, Physical Foundations of Cosmology}

\texttt{Mukhanov V.F., Feldman H.A., Brandenberger R.H., 1992,
PhR, 215, 203}

\texttt{Norsen T., 2017, Foundations of Quantum Mechanics: An Exploration of the Physical Meaning of Quantum Theory}

\texttt{Ocampo M.M., 2023, Explorando la transici\'on cu\'antico-cl\'asica de las perturbaciones primordiales durante la \'epoca inflacionaria del universo}

\texttt{Okon E., 2014, Revista mexicana de f\'isica E, 60, 130}

\texttt{Palermo O., et al., 2022, Eur. Phys. J. C, 82, 1146}

\texttt{Pearle P., 1976, Phys. Rev. D, 13, 857}

\texttt{Pearle P., 1989, Phys. Rev. A, 39, 2277}

\texttt{Pearle P., 2012, arXiv e-prints, arXiv:1209.5082}

\texttt{Penrose R., 1989, The emperor's new mind. Concerning computers, minds and laws of physics}

\texttt{Penrose R., 1996, Gen. Rel. Grav., 28, 581}

\texttt{Perez A., Sahlmann H., Sudarsky D., 2006, Classical and
Quantum Gravity, 23, 2317}

\texttt{Planck Collaboration, et al., 2020, A\&A, 641, A6}

\texttt{Sachs R.K., Wolfe A.M., 1967, ApJ, 147, 73}

\texttt{Schlosshauer-Selbach M., 2007, Decoherence and the
Quantum-To-Classical Transition}

\texttt{Sriramkumar L., 2009, arXiv e-prints, arXiv:0904.4584}

\texttt{Sudarsky D., 2011, Int. J. Mod. Phys. D, 20, 509}

\texttt{Zel'dovich Y.B., 1972, MNRAS, 160, 1P}

\end{document}